\documentclass{article}
\usepackage[T1]{fontenc}
\usepackage{spconf,amsmath,graphicx,hyperref}
\usepackage{cite}
\makeatletter
\newcommand{\bstctlcite}[1]{\@bsphack\if@filesw\immediate\write\@auxout{\string\citation{#1}}\fi\@esphack}
\makeatother
\usepackage{amssymb}
\usepackage{booktabs}
\usepackage{multirow}
\IfFileExists{kotex.sty}{\usepackage{kotex}}{}
\usepackage{balance}
\usepackage[table]{xcolor}

\definecolor{resultred}{RGB}{230,90,90}
\definecolor{resultblue}{RGB}{70,130,220}
\newcommand{\badcell}[2]{\cellcolor{resultred!#1}#2}
\newcommand{\goodcell}[2]{\cellcolor{resultblue!#1}#2}
\title{Joint Analysis of Latent Dimensionality and Frame Rate\\ in Continuous Audio Encoders}
\name{Kyudan Jung$^1$, Sehyun Lee$^1$, Song-ha Jo$^2$, Jaegul Choo$^1$, Sanghyuk Choi$^{3}$}
\address{$^1$KAIST AI, $^2$Seoul National University, $^3$NAVER Cloud}

\begin{document}

\maketitle
\bstctlcite{paper:ieee-style}

\begin{abstract}
Continuous audio encoders compress audio along feature and time axes through latent width and frame rate, but their joint effect on downstream performance remains unclear.
We train sixteen encoders spanning four widths and four frame rates, with downstream adapters and probes, using matched training protocols.
Despite generally improved reconstruction at larger widths, automatic speech recognition (ASR) and spoken question answering (SQA) favor moderate widths at higher rates, with the best observed widths shifting toward larger values under stronger temporal compression.
Frozen-model PCA interventions reveal distinct reconstruction and recognition sensitivities: removing the trailing half of the components substantially degrades ASR in high-rate 512-dimensional encoders with comparatively small reconstruction penalties, whereas 1024-dimensional encoders largely preserve both.
Yet the projected 1024-dimensional model underperforms unmodified narrower models on ASR at 12.5\,Hz.
These findings identify a width--rate interaction in downstream utility and suggest that how representations are organized during training matters beyond reconstruction fidelity and compressibility.
\end{abstract}

\vspace{-1pt}
\section{Introduction}
\label{sec:intro}

Speech language models and audio generation systems increasingly use neural codec representations~\cite{borsos2023audiolm,wang2023valle,defossez2024moshi}.
A continuous audio encoder converts audio into frame vectors, whose width $d$ and frame rate $f$ set the compression along the feature and time axes.
Lower frame rates shorten downstream sequences, motivating recent low-rate codecs~\cite{defossez2024moshi,ji2024wavtokenizer,xin2024bigcodec}.
Semantic encoders such as Whisper~\cite{radford2023robust}
lack waveform decoders, so reconstruction cannot be measured alongside downstream utility; we therefore study continuous codecs.

Codec settings are typically selected by reconstruction metrics, which improve with both width and frame rate.
Prior work varies latent width and frame rate one at a time, measuring reconstruction and TTS generation~\cite{longcat2026}, and explores frame rate selection for quantized speech QA~\cite{ye2026speechtext}.
Whether this trend extends to recognition models reading continuous representations remains unclear.

% Codec quality is commonly evaluated through reconstruction metrics.
% Wider vectors and higher frame rates generally improve reconstruction by providing more room for acoustic detail.
% However, our experiments show that accurate codec reconstruction does not guarantee strong downstream performance, and performance trends vary across tasks.
% Building on the semantic limitations of acoustic codecs~\cite{ye2024xcodec}, we examine how this gap depends on latent width and frame rate.

Our central hypothesis is that width and frame rate jointly shape how task-relevant information is retained and organized during training, a perspective motivated by the information bottleneck framework~\cite{tishby2000information}. An overly restrictive bottleneck may discard information required for the task.
An overly permissive one may preserve acoustic details without encouraging the encoder to organize task-relevant information for downstream use.
This interpretation follows the distinction between reconstruction objectives and representation quality~\cite{alemi2018fixing}.
The benefit of additional width should therefore depend on frame rate.
A moderate width may suffice with many frames, whereas sparse frames must summarize longer audio spans and may require more capacity.
We therefore expect the useful width to change with temporal compression.

We train all sixteen combinations of
$d\in\{128,256,512,$ $1024\}$ and
$f\in\{3.125,6.25,12.5,25\}$\,Hz using a common backbone and training recipe.
We evaluate reconstruction quality and downstream performance on automatic speech recognition (ASR), spoken question answering (SQA), and speaker identification.
The best observed ASR and SQA widths shift toward larger values as the frame rate decreases, suggesting that additional per-frame capacity becomes more beneficial under stronger temporal compression.

% Increasing width or frame rate does not consistently improve downstream performance.
% At 128--512 dimensions, reducing the rate from 25 to 12.5\,Hz leaves ASR performance nearly unchanged while improving SQA and speaker identification, with half as many audio frames.
% The best observed ASR and SQA settings use 256--512 dimensions at 12.5--25\,Hz, whereas both favor 1024 dimensions at 3.125\,Hz.
% Speaker identification follows a different pattern, indicating a task-dependent balance.

We further analyze these representations using PCA interventions without retraining.
At 1024 dimensions, removing the trailing half of the components largely preserves reconstruction and ASR performance, yet does not recover the ASR performance of unmodified narrower models at 12.5\,Hz.
This suggests that learning with a narrower bottleneck differs from removing dispensable directions afterward: the benefit of latent capacity may depend on how the bottleneck organizes information during training under a given frame rate.

Our contributions are as follows.
\begin{itemize}
\itemsep0pt
% \item We evaluate sixteen combinations of latent width and frame rate under a common codec training recipe.
% \item We show that downstream performance does not consistently follow reconstruction quality, and that the best observed width depends on frame rate and task.
% \item We use PCA interventions to reveal differing reconstruction and ASR sensitivities, supporting the role of learned information organization in downstream utility.
\item We train sixteen encoders and downstream adapters and probes, finding that reconstruction gains need not improve downstream performance.
\vspace{-1pt}
\item We identify a width--rate interaction: ASR and SQA favor moderate widths at higher frame rates and larger widths at lower rates.
\vspace{-1pt}
\item Frozen-model PCA reveals different reconstruction and ASR sensitivities to component removal.

\end{itemize}

\section{Methodology}
\label{sec:method}
%\vspace{-10pt}
\subsection{Architecture}
\label{sec:grid-method}
We study the two choices that directly define a continuous audio interface.
We train all sixteen combinations of $d \in \{128,256,512,1024\}$ and
$f \in \{3.125,6.25,12.5,25\}$\,Hz.
We start from the released DAC-VAE backbone~\cite{moviegen}, following DAC's convolutional design~\cite{kumar2024dac}. The front encoder and later decoder blocks remain frozen. For every model, we learn replacements for the final encoder block, its activation and projection, the VAE input/output projections, and the first decoder projection. The backbone maps 48-kHz audio to 25 frames per second.
For lower rates, we add $\log_2(25/f)$ learned stride-2 convolutional downsamplers before the bottleneck and corresponding transposed-convolution upsamplers after it. The decoder receives 25-Hz features, while each encoder is trained with its target-rate bottleneck rather than compressed post hoc.

\subsection{Training Objective and Dataset}
We optimize every cell with the same objective, 
\begin{equation}
\mathcal{L}=\lambda_{\mathrm{mel}}\mathcal{L}_{\mathrm{mel}}+
\lambda_{\mathrm{stft}}\mathcal{L}_{\mathrm{STFT}}+
\lambda_{\mathrm{kl}}\mathcal{L}_{\mathrm{KL}}+
\lambda_{\mathrm{gan}}\mathcal{L}_{\mathrm{GAN}}+
\lambda_{\mathrm{fm}}\mathcal{L}_{\mathrm{FM}}.
\end{equation}
We follow the DAC-VAE training recipe: $\mathcal{L}_{\mathrm{mel}}$ is a seven-scale mel-spectrogram loss,
$\mathcal{L}_{\mathrm{STFT}}$ is a multi-scale STFT loss, and
$\mathcal{L}_{\mathrm{FM}}$ is discriminator feature matching~\cite{kumar2024dac}.
We set $(\lambda_{\mathrm{mel}},\lambda_{\mathrm{stft}},\lambda_{\mathrm{kl}},
\lambda_{\mathrm{gan}},\lambda_{\mathrm{fm}})=(100,1,10^{-3},1,2)$.
We use a least-squares adversarial objective with five multi-period and three
multi-resolution discriminators. Adversarial training starts after five epochs.
The objective is shared across widths and rates.
All cells use the same mixed training data of speech, general sound, and music from LibriSpeech~\cite{panayotov2015librispeech}, WavCaps~\cite{mei2023wavcaps}, and MAESTRO~\cite{hawthorne2019maestro}. The implementation interleaves the sources so that all three modalities are included in a single batch. Each model is trained for 50 epochs with AdamW~\cite{loshchilov2019decoupledweightdecayregularization}, a learning rate of $3\times10^{-5}$, batch size 8 per worker, and 3,000 steps per epoch.

\section{Experiments}
\label{sec:setup}
We evaluate reconstruction quality and downstream performance across sixteen DAC-VAE-based codecs, then use PCA interventions to examine their reliance on latent directions.

\subsection{Reconstruction}
For reconstruction, we evaluate fifty 5.12-second segments from the LibriSpeech test set. We report wideband PESQ~\cite{rix2001pesq} at 16\,kHz and zero-mean SI-SDR~\cite{le2019sdr} at 48\,kHz. PESQ assesses perceptual speech quality, while SI-SDR measures waveform fidelity.

\subsection{Downstream Tasks}
\label{sec:downstream-method}
Following the frozen-representation evaluation approach of SUPERB~\cite{yang21c_interspeech}, we evaluate automatic speech recognition (ASR), spoken question answering (SQA), and speaker identification using task-specific adapters and probes.

\begin{table}[t]
\centering
%\vspace{-7pt}
\caption{Speech reconstruction results across all sixteen encoders.
PESQ and SI-SDR are higher when better. Bold indicates the best result in each row.}
\label{tab:reconstruction}
\resizebox{0.98\linewidth}{!}{%
\begin{tabular}{@{}r|cccc|cccc@{}}
\addlinespace[3pt]
\toprule
& \multicolumn{4}{c|}{PESQ $\uparrow$}
& \multicolumn{4}{c}{SI-SDR (dB) $\uparrow$} \\
Hz
& 128d & 256d & 512d & 1024d
& 128d & 256d & 512d & 1024d \\
\midrule

\addlinespace[3pt]
3.125
& 2.79
& 3.07
& \textbf{3.12}
& 3.09
& 0.19
& 0.38
& 0.43
& \textbf{0.76} \\

\addlinespace[1.5pt]
6.25
& 3.24
& 3.52
& 3.58
& \textbf{3.62}
& 1.80
& 4.13
& \textbf{4.42}
& 4.40 \\

\addlinespace[1.5pt]
12.5
& 3.61
& 3.83
& 3.79
& \textbf{4.41}
& 5.13
& 5.51
& 6.06
& \textbf{11.04} \\

\addlinespace[1.5pt]
25.0
& 4.43
& 4.53
& 4.53
& \textbf{4.54}
& 11.54
& 14.36
& 14.80
& \textbf{15.17} \\

\bottomrule
\end{tabular}}
\end{table}

\begin{table*}[t]
\centering
%\vspace{-7pt}
\caption{Downstream results for sixteen configurations.
Bold marks the best result per metric in each row.
Paired-bootstrap 95\% CIs have $\approx0.2$-pp half-widths
for WER differences and include zero for SQA differences
$<1$ pp; test-set sampling only.}
\vspace{5pt}
\label{tab:downstream}
\small
\setlength{\tabcolsep}{3.5pt}
\resizebox{0.95\linewidth}{!}{%
\begin{tabular}{@{}r|rrrr|rrrr|rrrr|rrrr@{}}
\toprule
& \multicolumn{4}{c|}{ASR WER (\%) $\downarrow$}
& \multicolumn{4}{c|}{SQA Cover-EM (\%) $\uparrow$}
& \multicolumn{4}{c|}{Spk-Vox Acc (\%) $\uparrow$}
& \multicolumn{4}{c}{Spk-Libri Acc (\%) $\uparrow$} \\
Hz
& 128d & 256d & 512d & 1024d
& 128d & 256d & 512d & 1024d
& 128d & 256d & 512d & 1024d
& 128d & 256d & 512d & 1024d \\
\midrule

\addlinespace[3pt]
3.125
& 8.67
& 7.82
& 7.66
& \textbf{7.62}
& 7.22
& 13.05
& 13.53
& \textbf{13.74}
& 19.12
& 22.36
& \textbf{23.43}
& 22.14
& 94.28
& \textbf{97.14}
& \textbf{97.14}
& 96.02 \\

\addlinespace[1pt]
6.25
& 4.76
& 4.47
& \textbf{4.37}
& 4.58
& 19.10
& 22.37
& 21.17
& \textbf{26.58}
& 27.94
& 30.03
& \textbf{30.50}
& 27.37
& 98.95
& \textbf{99.41}
& 99.13
& 99.09 \\

\addlinespace[1pt]
12.5
& 3.91
& \textbf{3.77}
& 3.83
& 4.22
& 32.64
& 33.39
& \textbf{34.58}
& 31.38
& 35.75
& 35.89
& \textbf{36.31}
& 23.63
& 99.58
& \textbf{99.83}
& 99.72
& 98.67 \\

\addlinespace[1pt]
25.0
& 3.89
& \textbf{3.83}
& 3.89
& 3.91
& 28.95
& \textbf{32.08}
& 29.67
& 31.62
& 25.04
& 25.12
& \textbf{26.09}
& 25.78
& 97.84
& 97.66
& 98.50
& \textbf{98.95} \\

\bottomrule
\end{tabular}}
\vspace{-8pt}
\end{table*}

For ASR and SQA, we attach a frozen \mbox{Qwen3.5-4B} language model~\cite{qwen3_5} to the frozen codec latent representations via a trainable adapter. We independently train a four-layer transformer adapter for each codec, with input normalization, eight attention heads, and a hidden dimension of 1024. We train only the adapter for 40,000 steps using AdamW with a peak learning rate of $10^{-4}$, weight decay 0.01, and an effective batch size of 64 sequences. The ASR corpus contains 15.5M utterances from MLS English, GigaSpeech, LibriTTS-R, and English VoxPopuli. We introduce SQA training data from InstructS2S-200K~\cite{fang2025llamaomni} around step 20,000. Audio is sampled at 48\,kHz, and rate-specific sequence-packing limits keep the number of utterances per step matched.

ASR is evaluated using a transcription prompt and WER on the full LibriSpeech \texttt{test-clean} set~\cite{panayotov2015librispeech}. For SQA, we evaluate the 17,943-item test split of \texttt{mistralai/}\allowbreak\texttt{triviaqa\_speech}~\cite{joshi2017triviaqa} using Cover Exact Match (Cover-EM), which checks if the reference answer is included in the generated response. WER measures transcription accuracy, while SQA assesses whether the representations support question answering.

For speaker identification, we concatenate the temporal mean and standard deviation of each utterance's frames, using statistics pooling as in x-vector systems~\cite{snyder2018x}, and train a single linear softmax layer. We evaluate this probe on VoxCeleb1~\cite{nagrani2017voxceleb} (1,211 speakers) and LibriSpeech \texttt{train-}\allowbreak\texttt{clean-}\allowbreak\texttt{100}~\cite{panayotov2015librispeech} (251 speakers). Classifiers are trained for up to 30 epochs with AdamW (learning rate $10^{-3}$, weight decay $10^{-4}$), selecting the best model based on validation accuracy.

\section{Results and Analysis}
\label{sec:results}
We first characterize reconstruction and the width--rate interaction in downstream tasks. We then use PCA to examine fixed-model reliance on latent directions and discuss the implications for the compression-pressure hypothesis.

\vspace{-2pt}
\subsection{Reconstruction degrades as frame rate decreases}
\label{sec:recon}
\vspace{-2pt}
Table~\ref{tab:reconstruction} shows improving reconstruction with temporal rate. Wider frames generally help, but no sharp fidelity boundary identifies a downstream operating point. Width and rate play different roles. Width enriches each vector, while rate supplies more observations in time. Additional width can compensate for some, but not all, effects of temporal compression.

\vspace{-2pt}
\subsection{Useful width depends on temporal rate}
\label{sec:downstream}
\vspace{-2pt}
In Table~\ref{tab:downstream}, reducing the rate from 25 to 12.5\,Hz changes WER only slightly at 128--512d, while SQA and speaker identification generally improve. The adapter between speech encoder and LLM processes half as many audio tokens in 12.5\,Hz and its audio self-attention score matrix has approximately one quarter as many entries. Downstream utility therefore need not increase with temporal resolution.

The width ranking changes as frames become sparse. At 12.5--25\,Hz, the best ASR and SQA point estimates occur at 256--512 dimensions. At 6.25\,Hz, ASR favors 512d and SQA favors 1024d; at 3.125\,Hz, both favor 1024d. These rankings suggest a shift toward wider frames under stronger temporal compression, not a statistically established unique optimum. Speaker identification does not follow the same shift, emphasizing that the useful operating region depends on the task.

This pattern is consistent with a capacity-allocation trade-off: moderate width may suffice when many frames are available, while sparser sequences may benefit from greater per-frame capacity.
Reconstruction and downstream utility peak in different parts of the grid: PESQ and SI-SDR are highest at 1024d and 25\,Hz, whereas all four downstream metrics have their best observed scores at 12.5\,Hz, where 1024d gives the best reconstruction but the weakest downstream results.

\begin{table}[t]
\centering
\vspace{-7pt}
\caption{Fixed-weight top-half PCA ablation.
``Half'' reports performance after retaining the leading half of the
PCA components. $\Delta$ denotes the relative performance change:
$100(\mathrm{WER}_{\rm Full}/\mathrm{WER}_{\rm Half}-1)$ for ASR and
$100(\mathrm{Acc}_{\rm Half}/\mathrm{Acc}_{\rm Full}-1)$ for speaker
identification. Darker red indicates a larger absolute change.}
\vspace{5pt}
\label{tab:pca-inference}
\resizebox{\linewidth}{!}{%
\begin{tabular}{@{}rl|rrrr|rrrr@{}}
\toprule
& & \multicolumn{4}{c|}{ASR WER (\%) $\downarrow$}
& \multicolumn{4}{c}{Spk-Vox Acc (\%) $\uparrow$} \\
Hz & Setting
& 128d & 256d & 512d & 1024d
& 128d & 256d & 512d & 1024d \\
\midrule

\addlinespace[3pt]
\multirow{2}{*}{3.125}
& Half
& 22.15 & 12.28 & 9.71 & \textbf{7.53}
& 16.87 & 20.98 & \textbf{22.37} & 21.49 \\
& $\Delta\%$
& \badcell{26}{$-60.8$}
& \badcell{17}{$-36.3$}
& \badcell{12}{$-21.2$}
& \badcell{4}{$+1.3$}
& \badcell{35}{$-11.8$}
& \badcell{20}{$-6.2$}
& \badcell{16}{$-4.5$}
& \badcell{12}{$-2.9$} \\

\addlinespace[3pt]
\multirow{2}{*}{6.25}
& Half
& 13.10 & 6.17 & 4.89 & \textbf{4.56}
& 25.24 & 28.61 & \textbf{29.02} & 26.38 \\
& $\Delta\%$
& \badcell{27}{$-63.6$}
& \badcell{14}{$-27.7$}
& \badcell{8}{$-10.6$}
& \badcell{4}{$+0.5$}
& \badcell{30}{$-9.7$}
& \badcell{16}{$-4.7$}
& \badcell{17}{$-4.9$}
& \badcell{14}{$-3.6$} \\

\addlinespace[3pt]
\multirow{2}{*}{12.5}
& Half
& 8.04 & 7.37 & 30.48 & \textbf{4.19}
& 33.06 & 34.17 & \textbf{34.98} & 22.97 \\
& $\Delta\%$
& \badcell{22}{$-51.4$}
& \badcell{21}{$-48.9$}
& \badcell{35}{$-87.4$}
& \badcell{4}{$+0.7$}
& \badcell{24}{$-7.5$}
& \badcell{17}{$-4.8$}
& \badcell{14}{$-3.7$}
& \badcell{11}{$-2.8$} \\

\addlinespace[3pt]
\multirow{2}{*}{25}
& Half
& 16.13 & 8.48 & 30.79 & \textbf{3.93}
& 23.82 & 24.13 & \textbf{25.32} & 25.23 \\
& $\Delta\%$
& \badcell{31}{$-75.9$}
& \badcell{23}{$-54.8$}
& \badcell{35}{$-87.4$}
& \badcell{4}{$-0.4$}
& \badcell{17}{$-4.9$}
& \badcell{14}{$-3.9$}
& \badcell{12}{$-2.9$}
& \badcell{10}{$-2.2$} \\

\bottomrule
\end{tabular}}
\vspace{-9pt}
\end{table}

\vspace{-2pt}
\subsection{Latent information analysis with PCA}
\label{sec:pca-speaker}

\begin{table*}[t]
\centering
%\vspace{-7pt}
\caption{Codec reconstruction after the ASR top-half PCA intervention.
``Half'' reports reconstruction performance after retaining the leading
half of the PCA components. $\Delta$ denotes the difference between
Half and the corresponding full-representation result reported in
Table~\ref{tab:reconstruction}. For each reconstruction metric, blue
intensity is normalized across all sixteen configurations, with darker
blue indicating better performance. Darker red indicates a larger
absolute degradation.}
\vspace{5pt}
\label{tab:pca-reconstruction-reference}
\small
\setlength{\tabcolsep}{3.5pt}
\resizebox{0.95\linewidth}{!}{%
\begin{tabular}{@{}r|rrrr|rrrr|rrrr|rrrr@{}}
\toprule
& \multicolumn{4}{c|}{PESQ Half $\uparrow$}
& \multicolumn{4}{c|}{PESQ $\Delta$}
& \multicolumn{4}{c|}{SI-SDR Half (dB) $\uparrow$}
& \multicolumn{4}{c}{SI-SDR $\Delta$ (dB)} \\
Hz
& 128d & 256d & 512d & 1024d
& 128d & 256d & 512d & 1024d
& 128d & 256d & 512d & 1024d
& 128d & 256d & 512d & 1024d \\
\midrule

\addlinespace[3pt]
3.125
& \goodcell{6}{1.98}
& \goodcell{12}{2.48}
& \goodcell{19}{3.09}
& \goodcell{19}{3.09}
& \badcell{14}{$-0.81$}
& \badcell{11}{$-0.59$}
& \badcell{4}{$-0.03$}
& \badcell{0}{$+0.00$}
& \goodcell{5}{$-6.51$}
& \goodcell{12}{$-1.42$}
& \goodcell{14}{0.30}
& \goodcell{15}{0.76}
& \badcell{28}{$-6.70$}
& \badcell{11}{$-1.80$}
& \badcell{4}{$-0.13$}
& \badcell{0}{$+0.00$} \\

\addlinespace[3pt]
6.25
& \goodcell{10}{2.27}
& \goodcell{16}{2.83}
& \goodcell{24}{3.55}
& \goodcell{25}{3.62}
& \badcell{16}{$-0.97$}
& \badcell{12}{$-0.69$}
& \badcell{4}{$-0.03$}
& \badcell{0}{$+0.00$}
& \goodcell{12}{$-1.52$}
& \goodcell{17}{2.49}
& \goodcell{20}{4.24}
& \goodcell{20}{4.40}
& \badcell{16}{$-3.32$}
& \badcell{10}{$-1.64$}
& \badcell{5}{$-0.18$}
& \badcell{0}{$+0.00$} \\

\addlinespace[3pt]
12.5
& \goodcell{12}{2.47}
& \goodcell{17}{2.89}
& \goodcell{26}{3.72}
& \goodcell{34}{4.41}
& \badcell{18}{$-1.14$}
& \badcell{15}{$-0.94$}
& \badcell{5}{$-0.07$}
& \badcell{0}{$+0.00$}
& \goodcell{16}{1.66}
& \goodcell{19}{3.68}
& \goodcell{22}{5.92}
& \goodcell{29}{11.04}
& \badcell{17}{$-3.47$}
& \badcell{11}{$-1.83$}
& \badcell{5}{$-0.14$}
& \badcell{0}{$+0.00$} \\

\addlinespace[3pt]
25.0
& \goodcell{5}{1.85}
& \goodcell{12}{2.50}
& \goodcell{33}{4.32}
& \goodcell{35}{\textbf{4.54}}
& \badcell{35}{$-2.58$}
& \badcell{28}{$-2.03$}
& \badcell{7}{$-0.21$}
& \badcell{0}{$+0.00$}
& \goodcell{18}{3.06}
& \goodcell{26}{8.60}
& \goodcell{33}{13.95}
& \goodcell{35}{\textbf{15.17}}
& \badcell{35}{$-8.48$}
& \badcell{25}{$-5.76$}
& \badcell{7}{$-0.85$}
& \badcell{0}{$+0.00$} \\

\bottomrule
\end{tabular}%
}
%\vspace{-10pt}
\end{table*}
\vspace{-2pt}
We further investigate why $1024$-dimensional representations sometimes underperform their $512$-dimensional counterparts. 
Because neural codecs are optimized for reconstruction, increasing the latent dimensionality may reduce the pressure to encode information compactly and may leave greater redundancy or unused capacity in the representation.
To examine downstream reliance on latent directions, we fit PCA on representations standardized using training-set statistics and retain the leading half of the components. We restore the original dimensionality and scale before feeding the projected representations into the frozen downstream models. For ASR, PCA is fit using latent frames extracted from 2,048 utterances in LibriSpeech \texttt{train-clean-100}~\cite{panayotov2015librispeech}. For speaker identification, PCA is fit on pooled training representations, again retaining half of the original feature dimensions.
Table~\ref{tab:pca-inference} shows that the ASR performance of the $1024$-dimensional models is largely preserved and in some cases even improves after the trailing components are removed.
This robustness suggests that the removed directions contribute little to ASR performance in the evaluated 1024-dimensional models. Narrower models generally show greater sensitivity to the same proportional reduction. However, these differences do not directly establish information density, since PCA ranks variance rather than task relevance and wider models retain more components.
For speaker identification, sensitivity to component removal also generally decreases as the latent width increases. 
Overall, these results indicate that robustness to PCA intervention is distinct from absolute downstream accuracy and varies substantially across downstream tasks.

\vspace{-2pt}
\subsection{Codec decoding after PCA}
\label{sec:pca-recon-method}
\vspace{-2pt}
To distinguish codec reconstruction sensitivity from downstream sensitivity, we feed the ASR PCA-modified latents into the corresponding frozen codec decoders for all sixteen settings. ``Half'' denotes reconstruction using latents that retain only the leading half of the PCA components, and $\Delta$ denotes the change in reconstruction performance relative to the original, unmodified latents.
We compute wideband PESQ at 16\,kHz and zero-mean SI-SDR at 48\,kHz using the same checkpoints, segments, and evaluation protocol as in Table~\ref{tab:reconstruction}.

\label{sec:pca-reconstruction}
Table~\ref{tab:pca-reconstruction-reference} shows that the mean reconstruction penalty decreases with width at every tested frame rate: 128d is most affected, followed by 256d and 512d, while 1024d is largely unaffected. 
This result strengthens the distinction between nominal capacity and directions needed by trained models. At 1024d, removing the trailing half of the standardized PCA components preserves both codec reconstruction and ASR performance, indicating a largely dispensable subspace for these evaluated interfaces. It does not mean that half of the information has been removed. PCA orders components by explained variance~\cite{jolliffe2016principal}, which does not directly measure task relevance. Moreover, the retained rank itself grows with width. Nor does post-hoc tolerance imply that training an encoder at half the width would yield the same representation. Nevertheless, these results show that preserving reconstruction quality and tolerating component removal are insufficient to explain downstream performance, suggesting that the benefit of latent capacity depends on how task-relevant information is organized under temporal compression.

More importantly, the reconstruction and ASR sensitivity rankings differ. High-rate 512d ASR is particularly unstable under the intervention, although its reconstruction penalty is much smaller than that of 128d or 256d. The codec decoder can therefore remain comparatively tolerant while the trained ASR interface fails. This further supports the mismatch between reconstruction quality and downstream performance.
This supports interpreting the large WER increase as more than a simple measure of broad waveform degradation.
However, the reconstruction test uses cropped inputs and aggregate acoustic metrics, so it cannot rule out loss of specific linguistic cues. The evidence motivates representation--model compatibility as an explanation to test, rather than equating either low reconstruction loss or high retained variance with downstream accessibility.

% \subsection{Compression pressure and downstream compatibility}
\subsection{Bottlenecks and Representation Learning}
\label{sec:analysis}

Reconstruction-oriented training makes information usable by the codec decoder, but not necessarily accessible to downstream models. Crucially, training with a narrower bottleneck differs from removing dispensable directions after training. At 1024d, top-half PCA largely preserves reconstruction and ASR performance, yet ASR performance remains below that of unmodified narrower models at 12.5\,Hz. Since PCA projection does not relearn features, this suggests that the bottleneck's benefit lies in how it shapes representations during training, beyond simply reducing excess dimensions.

The favorable bottleneck also depends on temporal compression: moderate widths perform well at higher frame rates, whereas wider latents become more beneficial as frames become sparse. Together, the grid results and PCA analysis support jointly selecting width and rate for each downstream task, with attention to how information is organized rather than reconstruction quality alone. PCA supports this interpretation, but establishing the compression-pressure mechanism requires controlled training experiments.

\section{Conclusion}
\label{sec:conclusion}

Across sixteen encoders under one recipe, reconstruction peaks at the widest,
highest-rate corner of the grid; none of the four downstream measures does. All four peak at 12.5\,Hz with 256--512 dimensions, at half the sequence length, where ASR matches its 25\,Hz accuracy and the other three improve. ASR and SQA additionally favor 1024 dimensions at the lowest frame rate.
PCA sharpens the split. Removing the trailing half of the components leaves 1024-dimensional encoders nearly unchanged in reconstruction
and ASR, yet severely degrades ASR in high-rate 512-dimensional encoders despite a small reconstruction penalty:
the decoder tolerates what the recognizer cannot. The projected 1024-dimensional model still trails unmodified narrower models at 12.5\,Hz, so post-hoc removal differs from training under a narrower bottleneck. What a latent preserves and what it makes accessible are set by width and rate together.

\label{PGstartrefs}
\balance

\bibliographystyle{IEEEtran}
\bibliography{custom}

\end{document}